\documentclass{mem2}
\usepackage{natbib}
\usepackage{txfonts}\usepackage{balance}
\usepackage{graphicx}
\usepackage[a4paper]{hyperref}
\idline{1}{1}
\begin{document}
\def\teff{$T\rm_{eff }$}
\def\kms{$\mathrm {km s}^{-1}$}

\title{Stellar evolution in real time:
Period variations in galactic RR Lyr stars}

   \subtitle{}

\author{
E.\,Poretti\inst{1,2},
J.F.\,Le Borgne\inst{2,3},
J.\,Vandenbroere\inst{2},
A.\,Paschke\inst{2,4},
A.\,Klotz\inst{5},
M.\,Bo\"er\inst{6},
Y.\,Damerdji\inst{6,7},
M.\,Martignoni\inst{2,4}, \and
F.\,Acerbi\inst{2}
          }

  \offprints{E. Poretti}

\institute{
INAF -- Osservatorio Astronomico di Brera, Via E. Bianchi 46,
I-23807 Merate, Italy.\\
\email{ennio.poretti@brera.inaf.it}
\and
GEOS (Groupe Europ\'een d'Observations Stellaires), 23 Parc de Levesville, 28300 Bailleau l'Ev\^eque, France
\and
Laboratoire d'Astrophysique, Observatoire Midi-Pyr\'en\'ees, Toulouse, France
\and
BAV, Munsterdamm 90, D-12169 Berlin, Germany
\and
CESR (CNRS-UPS), Observatoire Midi-Pyr\'en\'es, Toulouse, France
\and
Observatoire de Haute-Provence (CNRS/OAMP), France
\and
IAG, Universit\'e de Li\`ege,  B-4000 Li\`ege, Belgium
}

\authorrunning{Poretti et al.}

\titlerunning{Stellar evolution in real time}

\abstract{   
The times of maximum brightness collected in the GEOS RR Lyr database allowed
us to trace the period variations of a sample of 123 galactic RRab variables.
These data span a time baseline exceeding 100~years.
Clear evidence of period increases or decreases at constant rates has been
found, suggesting evolutionary effects. The observed rates are slightly 
larger than those predicted by theoretical models; moreover, there is an
unexpected large percentage of RRab stars showing a period decrease. The
new possibilities offered by the use of robotic telecopes (TAROTs, REM) 
and of data from satellite (CoRoT) are expected to speed up the project
to measure stellar evolution in real time.
}
\maketitle{}

\section{Introduction} 
Horizontal branch stars cross the instability strip in different
stages of their evolution; in these phases they become RR Lyr stars. 
The crossing
can take place in both directions; as a consequence, the periods
will be either increasing if the stars evolve from blue to red or
decreasing if they evolve from red to blue. Despite its importance
as a test for the stellar evolution theory, the observed rate of the period
changes is still an unknown quantity. 

The amateur/professional association GEOS (Groupe Europ\'een d'Observations Stellaires)
has built a database aiming to put together all possible RR Lyr light maximum times
published in the literature.
 The publications from the end of the XIX$^{\rm th}$
century up to today were scanned for this purpose and 
 recent observations from amateur astronomers of the European groups GEOS and BAV
(Bundesdeutsche Arbeitsgemeinschaft f\"ur Ver\"anderliche Sterne) fed the
database in a continuous way.
To date, it contains about 50000 maximum times from more than 3000 RR Lyr stars.
The GEOS database is freely accessible on the internet at
the address http://dbrr.ast.obs-mip.fr/. 
Therefore, this database is the adequate tool to search for secular period
variations in the galactic RR Lyr  stars, since  most of them
have been studied for tens of years, several of them since the end of the XIX$^{\rm th}$ century.
The first result of the project, {\it Stellar evolution in real time}, have
been presented by \citet{flb}. It is noteworthy that 
its outlines have been sketched during several GEOS meetings,
where the different knowledge of amateur and professional astronomers found
a very profitable synthesis.

\begin{figure}
\includegraphics[width=1.0\columnwidth,height=0.80\columnwidth]{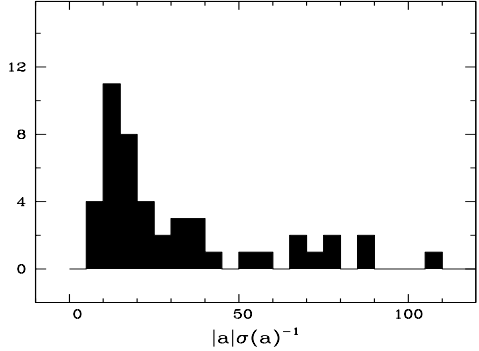}
\caption{\footnotesize Histogram of the ratios between the modulus of the $a$ coefficient and
its formal error.
}
\label{isto}
\end{figure}
\begin{figure*}
\begin{center}
\includegraphics[width=0.666666\columnwidth,height=0.6666\columnwidth]{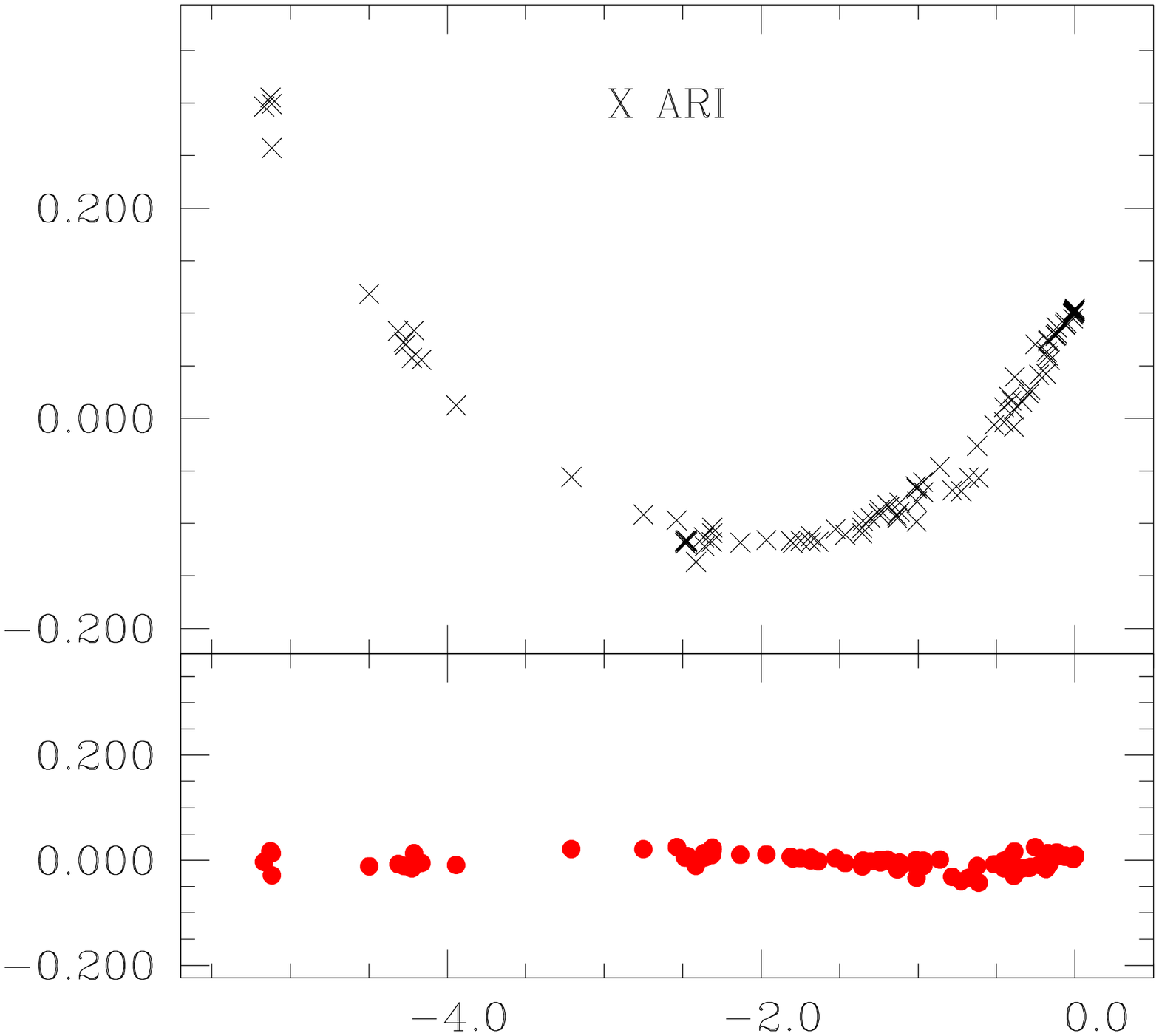}
\includegraphics[width=0.666666\columnwidth,height=0.6666\columnwidth]{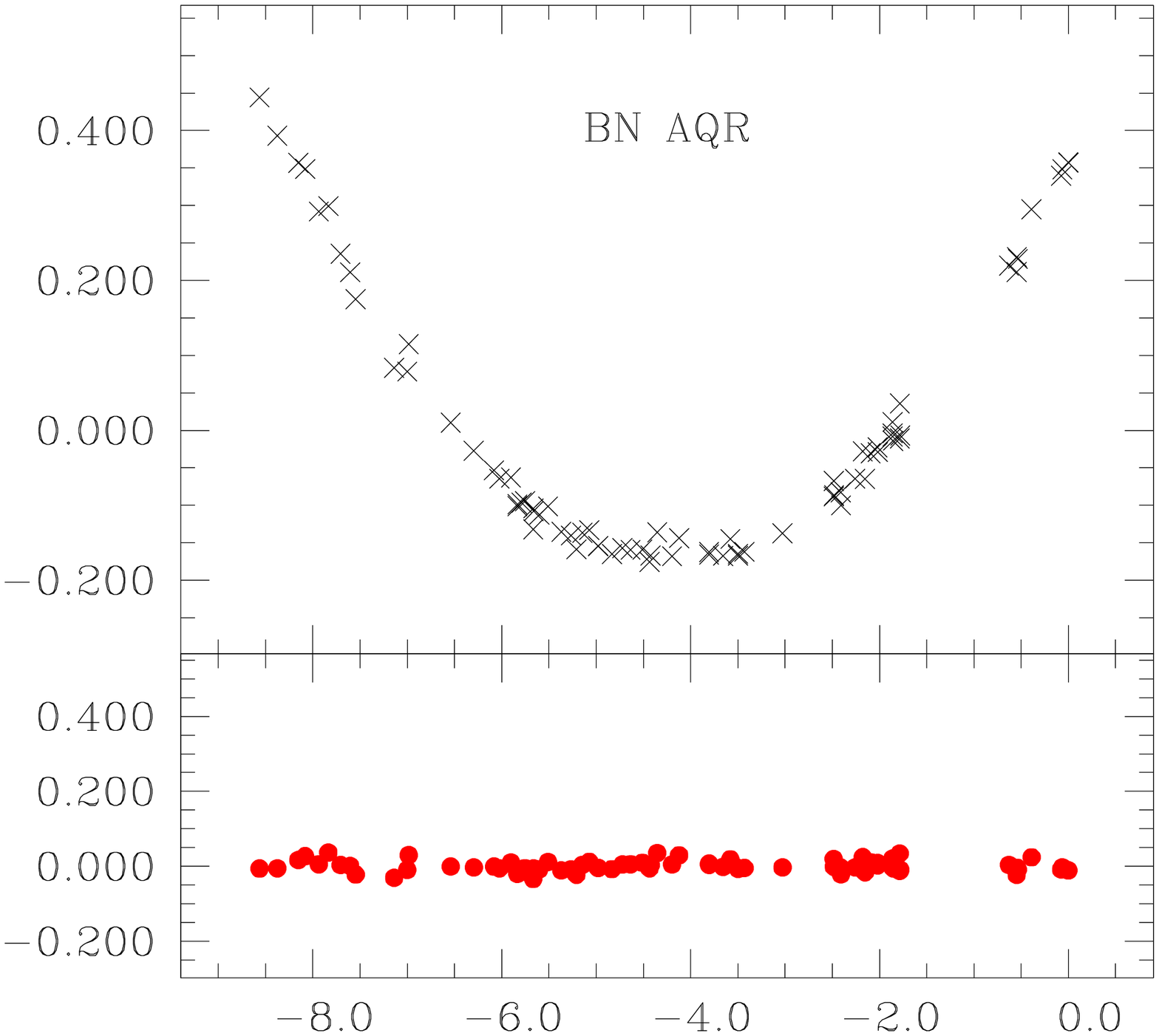}
\includegraphics[width=0.666666\columnwidth,height=0.6666\columnwidth]{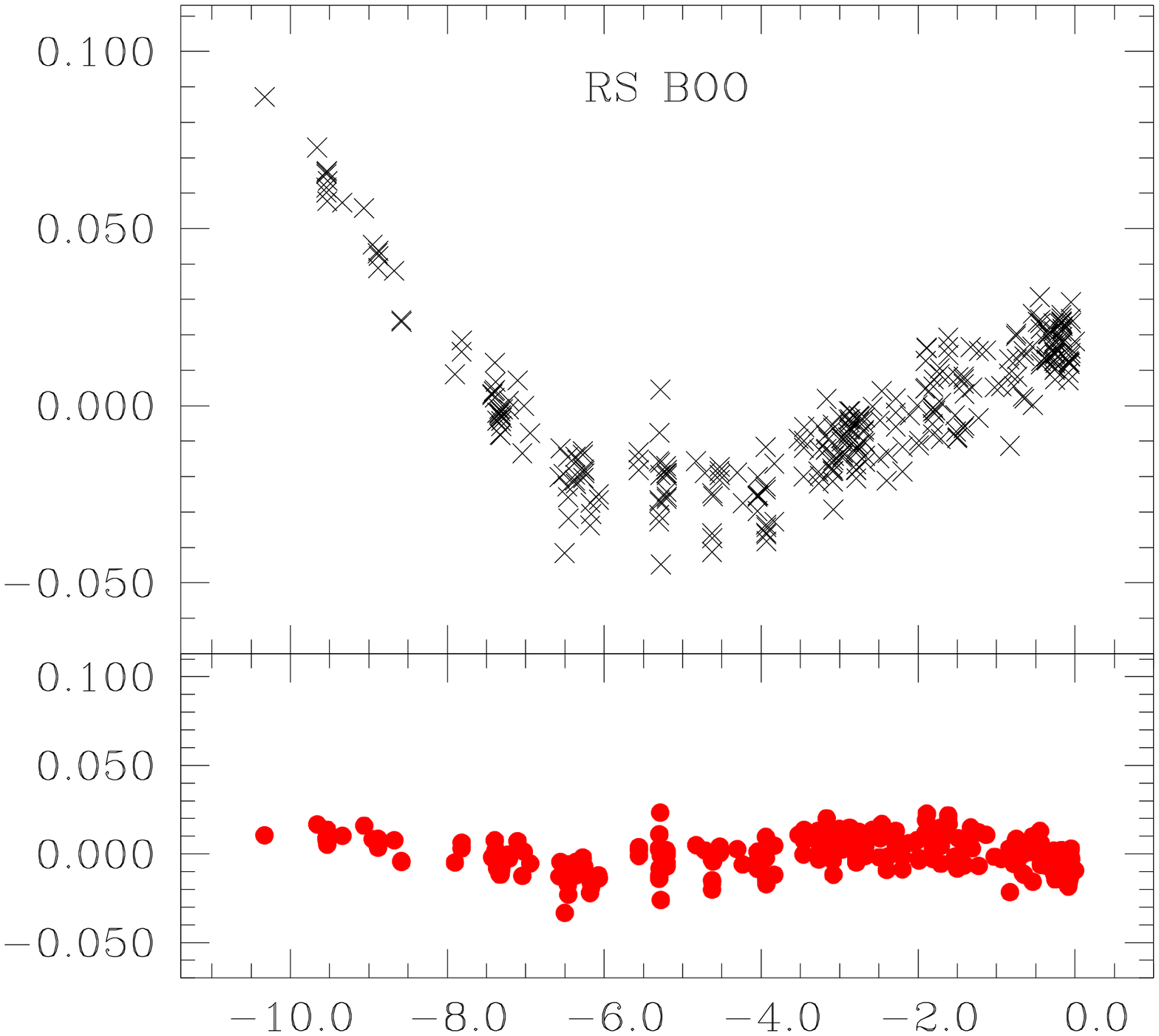}
\includegraphics[width=0.666666\columnwidth,height=0.6666\columnwidth]{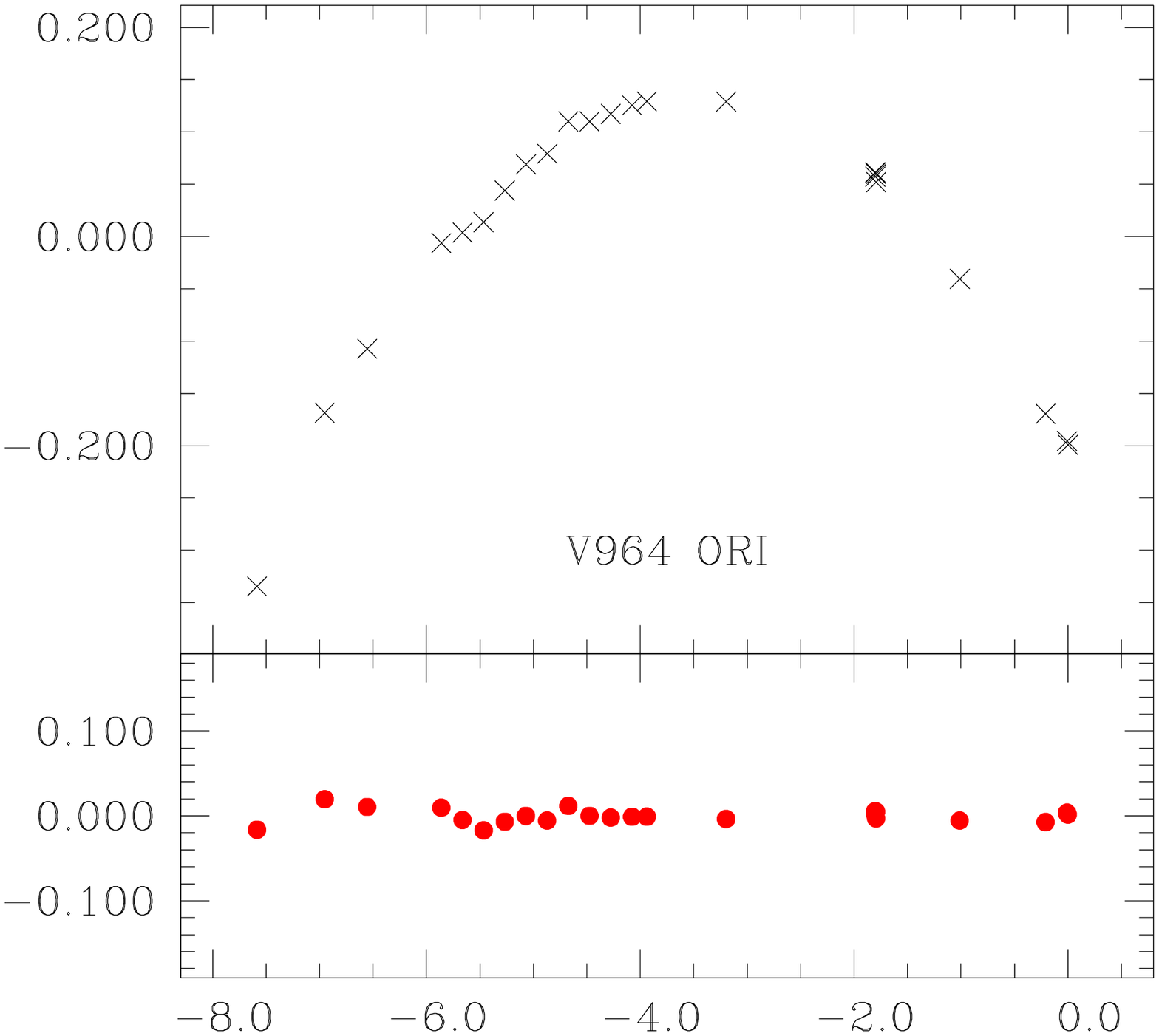}
\includegraphics[width=0.666666\columnwidth,height=0.6666\columnwidth]{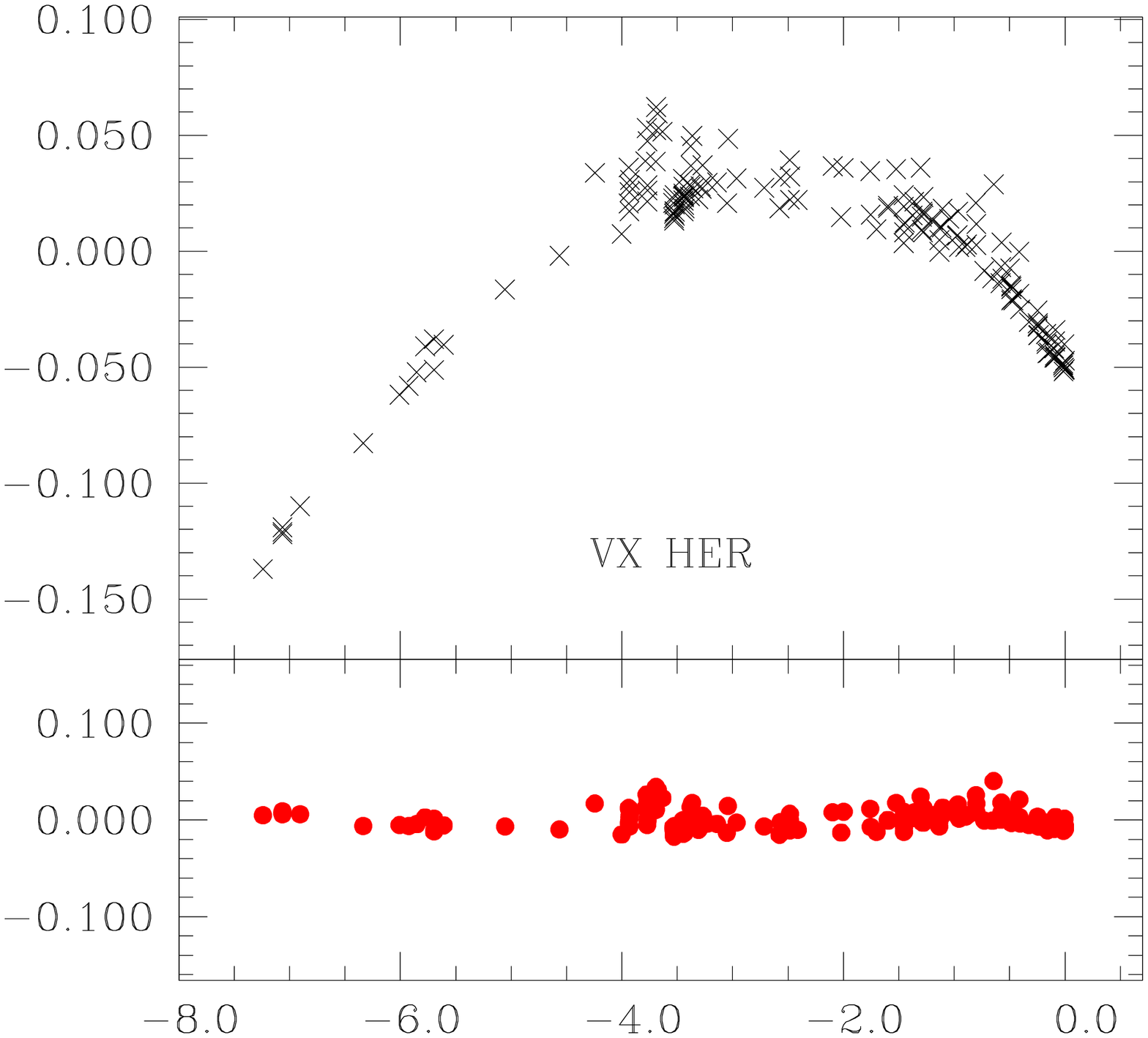}
\includegraphics[width=0.666666\columnwidth,height=0.6666\columnwidth]{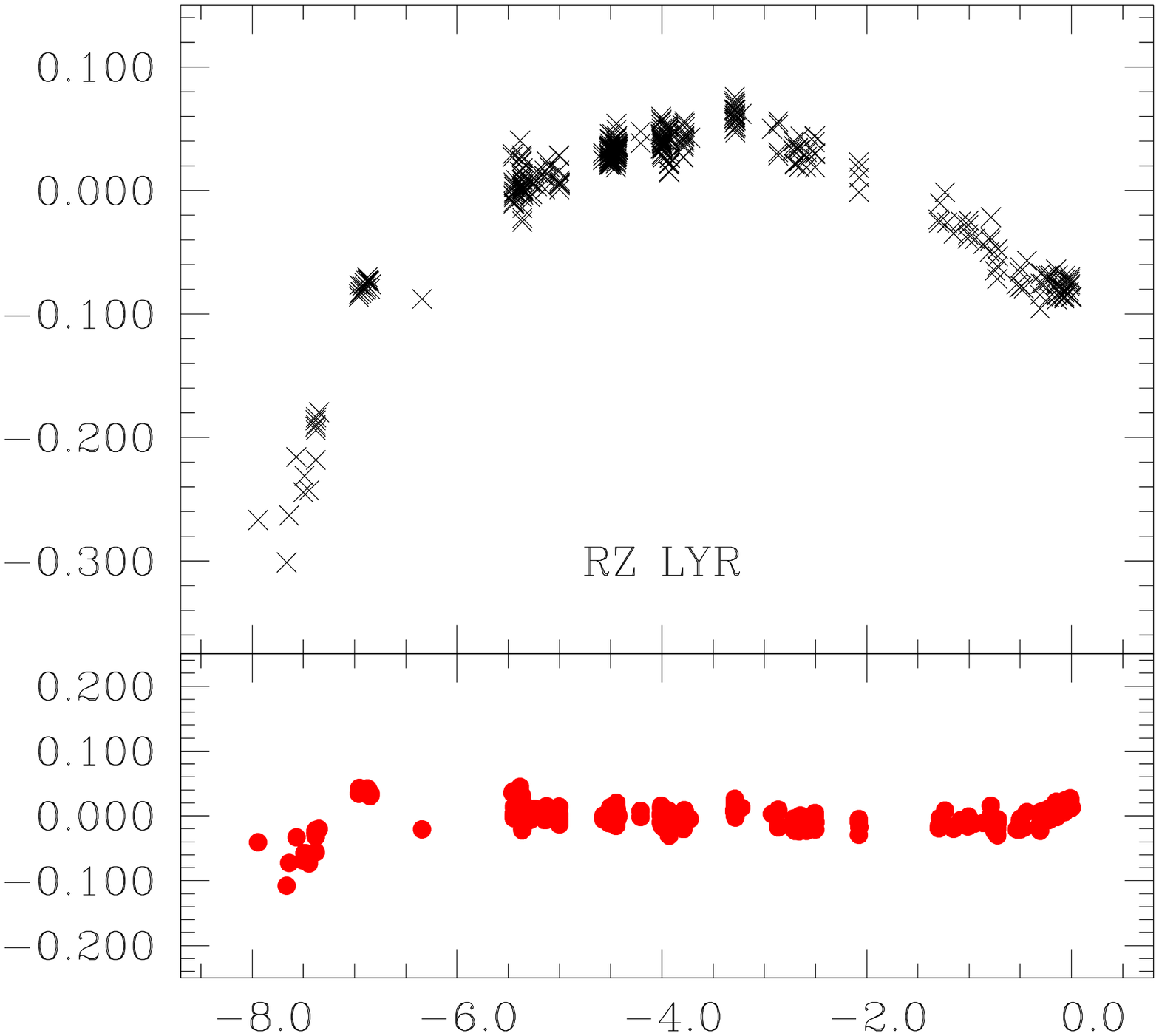}
\caption{\footnotesize For each star the upper panel shows the O--C values as
obtained from linear elements, the lower panel from a parabolic fit.
Measure units are elapsed cycles (E/10000) on the $x$--axis and days on the $y$-axis.}
\label{oc}
\end{center}
\end{figure*}

\section{The methodological approach: some insights}

To calculate the rates of the period changes, the times of maximum
brightness have been fitted  by means of  the linear elements
\begin{equation}
C = T_0 + P\cdot E
\end{equation}
where $P$ is the period (in days) and $E$ is the cycle counter.
In several cases the linear fit was not able to reproduce the observed
times of maximum brightness and a parabolic trend appeared in the O--C
(observed minus calculated values) plots. Therefore, the maxima were
fitted with a second--order polynomial 
\begin{equation}
C = T_0 + P_0\cdot E + a\cdot E^2
\end{equation}
where $P_0$ is the period at  epoch $T_0$.
The value of the $a$ coefficient is related to the rate of the period change
by means of 
\begin{equation}
dP/dt = 2\cdot a/ \langle P \rangle
\end{equation}
where $dP/dt$ is measured in [d/d] and  $\langle P \rangle$ is the average period, i.e., the ratio between
the elapsed time 
and the elapsed cycles. 
Theoretical models often refer to the rate of period changes using
the $\beta$ parameter, expressed in  [d~Myr$^{-1}$], since it directly 
indicates how much the period change in 10$^6$~years. 

Though the times of maximum brightness have
been collected by using very different techniques (visual, photographic, photolectric,
CCD), the very long time baseline allowed us to trace the period variation in a clear way.
We extracted from the GEOS database the RRab stars for which the times of maximum
brightness span more than 50~years. Fifty--four stars show a constant period,
while 27 show an increasing period and 21 a decreasing one. The remaining 21 RRab variables
show erratic changes. 
Figure~\ref{isto} shows the histogram of the ratios between the modulus of the $a$ coefficient
and its formal error for the 48 stars showing a linearly variable period. The ratios are
 always greater than 5.0 and are greater than 10 in all cases except four.
Therefore, the parabolic trends are very well defined and they 
bear witness of the reliability of the results. 

\begin{figure}
\includegraphics[width=1.00\columnwidth,height=0.40\columnwidth]{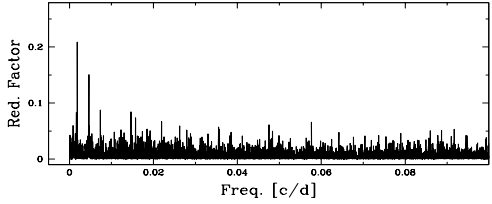}
\caption{\footnotesize RS Boo: power spectrum of the residual O--C values after having fitted
the maxima with a third--order polynomial. The highest peak is at 0.0019~d$^{-1}$ and it 
corresponds to the known period of the Blazhko effect.}
\label{rsboo}
\end{figure}

Figure~\ref{oc} illustrates some examples of the O--C behaviours.
Among the stars showing a period increase (top row), BN Aqr is one of the most evident cases: 
the parabolic trend left by the linear elements  is very well defined and
the two branches are symmetrical. No improvement has been obtained by introducing a
third order term; as a matter of fact, the residual O--C values are very flat around
the zero value. The O--C plot from the linear fit of the X Ari maxima is slightly different: 
the ascending branch seems
to be steeper than the descending one. As a consequence, the residual O--C values 
show an apparent oscillation. We carefully investigated
the nature of this oscillation, visible also in other cases. 
The related periodicities are often
about half the time coverage and their amplitudes are quite small. Therefore, we can
infer that the parabolic fit
is not perfect and that some systematic residuals were left. The maxima of
RS Boo shows this effect in a more pronounced way.
In some cases we improved the O--C fitting by
calculating a third--order least--squares fit, but in three cases only
we obtained a significant third--order term.
We note that no appreciable change is derived for the value of the 
$a$ coefficient when performing a third--order fit. 
The case of RS Boo is still more intriguing, since the residual O--C values
show a scatter on another timescale, shorter than the two previous ones.
As a matter of fact, when performing the frequency analysis of the residual
O--C values (i.e., the  prewhitened data after the third--order fit)
we were able to detect  a peak corresponding to at 534~d (Fig.~\ref{rsboo}), 
practically coincident with  the value given by  \citet{oost} for the
Blazhko period, i.e., 533~d.  The connection between period changes
and the Blazhko effect is still unclear and deserves more attention in
the future.

Among the stars showing a period decrease (bottom row), V964 Ori shows the  clearest pattern,
in spite of the small number of points. Asymmetrical O--C plots are also common, perhaps
more common than in the case of period increases, and VX Her constitutes a good example.
As in the case of RS Boo, also AH Cam shows a Blazhko effect, but on a very short
timescale (only 11.18~d, \citealt{smith}): it is notewothy that also in this case
the times of maximum brightness are able to determine the correct Blazhko period 
(see Fig.~6 in \citealt{flb}).


\section{Scientific conclusions}
\begin{figure}
\includegraphics[width=1.00\columnwidth,height=1.00\columnwidth]{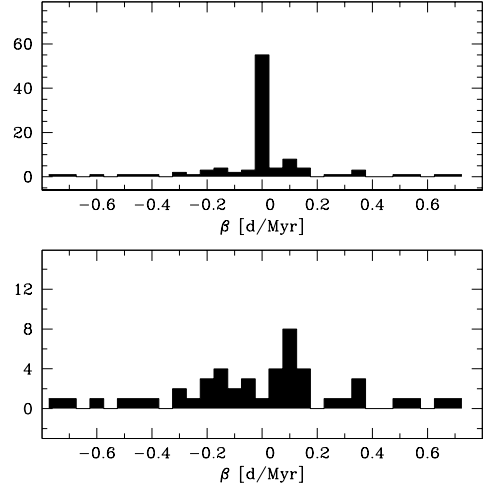}
\caption{\footnotesize Histograms of the distributions of the $\beta$ values
of the whole sample (upper panel) and restricted to the stars shown an 
appreciable linearly variable period (lower panel). The observed $\beta$ values are
slightly larger than those predicted by the evolutionary models.
}
\label{istobeta}
\end{figure}

The most remarkable result is the detection of a large number of O--C plots showing
a parabolic pattern when linear elements are used. The cause of this pattern is
a regular period variation, having the same order of magnitude of the
expected evolutionary changes.
On the other hand, in 54 cases (i.e., a number of cases similar to that of RR Lyr
stars showing period variations) we did not detect a significant period variation and
linear elements satisfy the available times of maximum brightness.
The upper panel in Fig.~\ref{istobeta} shows the distribution of the $\beta$ values 
in the sample of the galactic RRab stars.
The central peak is given by the stars showing a constant period, but
the number of stars both in the negative and in the positive planes
are relevant.

Indeed, the comparison between RRab stars showing negative and positive rates yields us the most
important scientific feedbacks.
They  can be evaluated in an easier way by removing the central peak (lower panel in Fig.~\ref{istobeta}).
It is quite evident that the blueward evolution
(decreasing periods, negative $\beta$ values; 21 cases) is as common as the redward one 
(increasing periods, positive $\beta$ values; 27 cases).  
By analyzing and comparing these two samples,
\citet{flb}  could stress some well--established observational facts:
\begin{enumerate}
\item RRab stars experiencing blueward evolution are quite common, only slightly less than
RRab stars experiencing redward evolution. The period
range covered by the two groups are very similar and the average and median periods
are nearly coincident;
\item the absolute values of period changes are larger than expected. The theoretical
 evolutionary rates calculated by \citet{lee} predict $\beta$ values smaller than 0.30
for almost all stars. $\beta$ values greater than 0.30 are  reported in exceptional
cases. Since in our sample we have 18 stars with $0.00<\beta<0.30$ and 9 stars with 
$\beta>$0.30, the 2:1 ratio seems too small to be caused by the exceptional conditions.
\item Blazhko effect is often superimposed on secular changes, but the monotonic
trend due to evolutionary variations still remains visible;
\end{enumerate}
In particular, the rates of the period changes are difficult to associate with 
canonical horizontal--branch evolution. The large observed $\beta$ values could be
ascribed to a sort of ``noise" superimposed on the canonical evolutionary tracks,
irrespective of where the star is in the course of its horizontal--branch evolution.
However, this is still surprising since the main sources of this ``noise" had
been invoked to act toward the exhaustion of the helium in the core, but not before
\citep{sweig}.
 At this purpose, we also note that there is a numerous group of RRab stars showing
erratic O--C behaviours, with abrupt or
 continuous period variations, never monotonic. The regular variations caused by light--time effect
(and hence duplicity),
 often invoked to explain large O--C excursions, are not convincingly observed in any
case (except maybe for RZ Cet)  of our sample.  
                                                                                                                                 
\section{The future}
As a general conclusion about the comparison between our observational results and the
theoretical predictions, we claim that there is a very powerful feedback between the two
approaches.
In particular, theoretical investigations should take into account that 
many RR Lyr stars are showing a 
blueward evolution. The theoretical models should also match the observed $\beta$ values in
a more satisfactory way, as these seem to be
higher than expected, both for redward and blueward evolutions.

To rapidly progress in the project {\it Stellar evolution in real time}
we need to accumulate much more accurate times of maximum brigthness. In such a case
the parabolic trend will be evidenced over a time baseline much shorter than the 110 yr interval
used for this first approach. The robotic telescopes 
TAROTs ({\it T\'elescope \`a Action Rapide pour les Objets
Transitoires} or {\it Rapid Action Telescope for Transient Objects};
\citealt{tarot}), operating
at La Silla (Chile) and at Calern Observatory (France), are currently used
to monitor RRab stars brighter than $V$=13.0.  They are providing very homogenous
list of CCD maxima (e.g., \citealt{LeBorgne07}) 
characterized by  small errorbars ($\pm$0.003~d on average). 

Another critical item is to investigate the interplay between Blazhko periods and
evolutionary changes, which can be evidenced only by means of dense lists of
maxima over one or two decades. 
We also have to note that the driving mechanism of the Blazhko
phenomenon needs some clarification, since the two competitive explanations,
the magnetic and the resonance models, are both still plausible. 
To pursue both goals in a specific way, we started a survey of few 
Blazhko RRab stars by using the robotic REM ({\it Rapid Eye Movement},
\citealt{rem}) telescope located at La Silla, Chile. 

Moreover, a substantial improvement in the study of the light curves of  RR Lyr variables could
be obtained by a continuous monitoring, without the seasonal  or the night--day alternances. 
Such a possibility was a dream until a few years
ago. After the successful launch on December 27th, 2006,  the satellite CoRoT 
({\it COnvection, ROtation and planetary Transits})
is now transforming the dream into reality. Staring on the same field for 150~d, 
CoRoT will monitor 10 bright
stars in the asteroseismologic channel and 12000--15000 in the exoplanetary one.
In the latter channel, the monitoring of some classical variables can be proposed 
in the framework of the ``Additional Programs" and a specific one on RR Lyr variables
has been approved (P.I.: M.~Chadid, Nice University; some of the authors are CoIs).
In such a context, the monitoring of three RR Lyr variables 
(V1127 Aql, V1220 Aql, and NSVS~12568727) has been proposed for the two first
CoRoT runs with the goal to investigate very subtle 
cycle--to--cycle variations (and hence large or small Blazhko effects, if any).
\begin{acknowledgements}
The authors wish to thank M.~Catelan for the very useful discussions during
the meeting and for  further comments.
EP acknowledges financial support from the Italian ESS project, contract ASI/INAF/I/015/07/0,
WP 3170.
\end{acknowledgements}

\end {document}